\documentstyle[preprint,aps]{revtex}
\input psfig
\begin{document}
\draft

\title{Intermittency, chaos and singular fluctuations in the mixed
 Obukhov-Novikov shell model of turbulence}
\author{Thierry Dombre and Jean-Louis Gilson}
\address{Centre de Recherches sur les Tr\`es Basses Temp\'eratures, CNRS,
BP166, 38042 Grenoble Cedex~9, France}

\maketitle
\begin{abstract}
The multiscaling properties of the mixed Obukhov-Novikov shell model
of turbulence are investigated numerically and compared with those of the
complex GOY model, mostly studied in the recent years. Two types of
generic singular fluctuations are identified~: first, self-similar
 solutions propagating from large to small scales and building up
intermittency, second, complex time singularities inhibiting the
cascade and promoting chaos. A simple and robust method is proposed to
track these objects. It is shown that the scaling exponent of self-similar
solutions selected by the dynamics is compatible with large order
statistics whenever it departs enough from the Kolmogorov value. Complex time
singularities on the other hand get trapped on the last shells, when the
 proportion of Novikov interactions exceeds a critical value which is
argued to mark the boundary between chaotic and regular dynamics
in the limit of infinite Reynolds number.
\end{abstract}
\pacs{PACS numbers 03.40.G, 47.27}

\section{Introduction}
The shell models of turbulence have recently attracted a lot of interest
as a useful tool for mimicking the Navier-Stokes dynamics. In the simplest
scalar models, one places velocity variables on a one-dimensional array
of wavevectors of the form $k_{n}=k_{0} Q^{n}$, where the integer $n$ labels
the shell from 0 to, ideally, $+\infty$ and $Q$ is a scale parameter fixing the
 step of the cascade. The time evolution of shell-velocities is governed by
ordinary differential equations with quadratic non-linearities, whose strength
grows like $k_{n}$, deterministic forcing at large scales and viscous
dissipation at small ones. The couplings between shells are usually local
and chosen in such a way that the total kinetic energy is conserved in the
absence of forcing and viscous effects. These hydrodynamic systems display
strong departure from the naive scaling expected on the basis of Kolmogorov-
like dimensional analysis, which shows up in particular in the higher order
moments of velocity. Following the seminal work of Okhitani and Yamada
\cite{YO,OY}, Jensen, Paladin and Vulpiani \cite{JPV} found in a particular
shell model, nowadays referred to as the GOY model, multiscaling
properties very close to those of real turbulent flows. Most subsequent
studies in this field have therefore concentrated on the GOY model and
important progress was made towards a deeper understanding of its behaviour
in recent publications \cite{BBP,BLLP,KLWB} (we shall go back to some of these
 results in the bulk of the paper).\\
The GOY model uses complex velocity variables and interactions among all triads
made up of three different neighbouring shells. One may wonder whether these
two features are necessary to produce "good" chaotic properties. In order
to clarify this question, we report in this paper a mostly numerical
investigation of the scaling properties of a simpler class of shell models,
which results from the linear superposition of two chains introduced in the
early $70'$s by Obukhov \cite{O} and Novikov \cite{DN} (for an historical
insight into the field and a comprehensive review of the huge russian
litterature
 concerned with cascade-like systems under various disguises, see
for instance \cite{GGO}). In the Obukhov-Novikov model, hereafter referred to
as
the ON model, the velocity variables are real and interacting triads involve
only two neighbouring shells. The structure of non-linearities depends as
in the GOY model on a single parameter (together with the scale
parameter $Q$), which fixes in that case the relative proportion of the two
 basic chains. Both models exhibit qualitatively the same
phenomenology. When the proportion of Novikov-like interactions
(favouring the transfer of energy towards small scales) is high enough, the
 system relaxes to a time independent state with Kolmogorov scaling properties.
As Obukhov-like interactions (favouring on the contrary the backflow of
energy towards large scales) take over, the system transits through a
 Ruelle-Takens scenario into a chaotic state with stochastic fluctuations.
There is clearly multifractality close to the transition, even if it looks
less pronounced than the one observed in the GOY model for usual values
of parameters.\\
We switch to more deterministic concerns in the second part of this paper,
which aims at characterizing singular fluctuations able to form in the
ON model or more generally in any one-dimensional shell model. We shall argue
that self-similar or soliton-like solutions of the equations of motion in the
inertial range are the building blocks of intermittency, while movable
singularities occuring at complex times induce chaos by inhibiting the
energy cascade. We are not aware of any previous study of the structure of
complex time singularities in shell models. In contrast, self-similar solutions
were already considered within the context of the ON model by Siggia \cite{S78}
and later on in more details by Nakano \cite{N88}. They are curiously
absent of more recent works. We propose here an efficient method for
identifying such solutions without any {\em a priori}
assumption on their shape. We find that the set of dynamically accessible
self-similar solutions is in fact limited to one single object (as Nakano's
results suggested it). This proves that multiscaling properties should not be
 ascribed to the existence of a large manifold of singular behaviours. The
 exponent $z$ controlling the multiplicative growth of these particular
 solutions, accounts in a satisfying way for the asymptotic scaling properties
 of high order velocity or energy transfer moments. It is easy to extend this
 analysis to the GOY model, where basically the same conclusions concerning the
 unicity of solutions can be drawn. However self-similar solutions in that case
are very mild  and do not seem to play a major role in the statistics at high
 orders. \\
The paper is organized as follows : in Section 2 we specify the conventions
used in our computations for normalizing variables and parameters and
describe some general properties of the ON model. Section 3 presents
statistical results obtained from numerics. The emphasis is put
on scaling exponents of the moments of energy transfer and their evolution with
 the relative proportion of Obukhov and Novikov interactions. Our goal here is
 to provide the reader with data and facts, disentangled from any theoretical
 interpretation. An attempt of comparison with the GOY model is made.
Section 4 is devoted to the hunt for self-similar solutions and a
 confrontation of their scaling properties with the statistics of the model at
 large orders. Complex time singularities are introduced and studied in Section
5, while perspectives and conclusions are briefly outlined in Section 6.

\section{General properties of the model}
\subsection{Definitions and basic considerations}
As already said in the Introduction, scalar shell models define a velocity
variable $u_{n}$, real or complex, on a one-dimensional array of wave-vectors
$k_{n}=k_{0}Q^{n}$ where the integer $n$ runs from $0$ to $N$. In most of the
 paper we shall restrict ourselves to the
case of real variables. It simplifies notations to consider that
the $u_{n}$ form a ($N+1$)-dimensional vector $\vec u$. The equation of motion
then takes the following form
\begin{equation}
\frac{d}{dt}\vec{u} =\vec{N}[\vec{u}]+\vec{F}-\vec{D}
\label{equmod1}
\end{equation}
where the three vectors $\vec N$, $\vec F$, $\vec D$ embody respectively the
 non-linearities, the external forcing and the dissipation. We only
considered a deterministic forcing acting on the $\mbox{zero}^{th}$ shell
and usual viscous dissipation, which means
\begin{equation}
F_{n}=f\delta _{n,0}\;\;\;,\;\;\; D_{n}=\nu k_{n}^{2}u_{n}
\end{equation}
where $\nu$ is the kinematic viscosity. The nonlinear kernel $\vec N$ is
quadratic in the $u_{n}$, with a coupling constant growing like $k_{n}$ in
order to reproduce the hierarchy of characteristic times of the Navier-Stokes
dynamics. It must also conserve the total kinetic energy
$E=\displaystyle{1\over 2}\sum_{n=0}^{N} u_{n}^{2}$. If interactions
between shells (which are always supposed to be local) do not extend beyond
nearest neighbours, the most general expression for the $n^{th}$ component of
 $\vec N$ is
\begin{equation}
N_{n}[\vec u] = \alpha Q^{2\over 3} [k_{n}u_{n-1}u_{n}-k_{n+1}u_{n+1}^2]
+\beta [k_{n}u_{n-1}^2-k_{n+1}u_{n}u_{n+1}]
\end{equation}
(this formula remains valid on the two boundaries $n=0$ and $n=N$,
provided  $u_{-1}=u_{N+1}=0$ is assumed).\\
The model appears like the linear superposition of the Obukhov-Gledzer (OG)
and Novikov-Desniansky (ND) chains, with respective weights $\alpha Q^{2/3}$
and $\beta$. We shall assume $\alpha,\beta >0$  and, without loss
of generality, $\alpha +\beta =1$. Since, on the average, the $u_{n}$
decrease like $k_{n}^{-1/3}$ according to Kolmogorov-scaling, it is convenient
to introduce a new set of variables $\phi_{n}$ by the relation
\begin{equation}
u_{n}=Q^{-{n\over 3}} \phi_{n}
\label{defphi}
\end{equation}
%The scaling properties of $\phi$ will give, if any, the deviations from K41.
The equations for $\vec \phi$ read
\begin{equation}
\frac{d}{dt}\vec{\phi} = k_{0} Q^{2\over 3} \vec{N}[\vec{\phi}]+\vec{F}-\vec{D}
\end{equation}
where the expression of the $n^{th}$ component of $\vec N$ is now
\begin{equation}
N_{n}[\vec{\phi}]=Q^{2n\over 3}[(\alpha \phi_{n-1}\phi_{n}+\beta \phi_{n-1}^2)
 - (\alpha \phi_{n+1}^2+\beta \phi_{n}\phi_{n+1})]
\label{nlphi_n}
\end{equation}
We still have the freedom to set to unity the forcing amplitude and the
coefficient in front of $\vec N$ by non-dimensionalizing in the proper way
time and velocities. The final form of the equations (as they were used in
the numerical investigations reported in Section 3) is
\begin{equation}
\frac{d}{dt} \phi_{n} =N_{n}[\vec{\phi}] +\delta_{n,0}-
{1\over R} Q^{2n} \phi_{n}
\label{equphi_n}
\end{equation}
The Reynolds number $R$ has been defined as
$\displaystyle{R={1\over \nu} \sqrt{{f Q^{2\over 3}}\over k_{0}^{3}}}$.
 Equations (\ref{nlphi_n}) and (\ref{equphi_n})
make energy conservation quite obvious. Indeed, in the limit $R=+\infty$ and
for $n\geq 1$, the energy $E_{n}=\displaystyle{u_{n}^{2}\over 2}=
\displaystyle{{1\over 2}{\phi_{n}^{2} Q^{-{2n\over 3}}}}$ carried by
the $n^{th}$ shell obeys the equation
\begin{equation}
\frac{d}{dt} E_{n}= \epsilon_{n}-\epsilon_{n+1}
\end{equation}
where
\begin{equation}
\epsilon_{n}= \phi_{n-1} \phi_{n} (\alpha \phi_{n} +\beta \phi_{n-1})
\label{flux_n}
\end{equation}
is the energy flux from the $(n-1)^{th}$ to the $n^{th}$ shell. Kolmogorov
scaling corresponds to $\phi_{n}=C^{te}$ or more fundamentally to
$\epsilon_{n}=C^{te}$ throughout the cascade.\\

The physics of the model, as defined by equation (\ref{equphi_n}), depends
on three parameters, namely : the step of the cascade Q, the
 proportion of Novikov interactions $\beta$, and the Reynolds number $R$.
The number of shells will not matter, provided the truncation is done far
beyond the Kolmogorov dissipative scale, where viscous effects become of the
same order as inertial ones. Assuming $\phi_{n}=O(1)$, the index $N_{d}$ of
the dissipative shell is given by the condition $\displaystyle{{Q^{2N_{d}}
\over R}\sim Q^{2N_{d}\over 3}}$. One should however pay attention to the
fact that the stronger the fluctuation, the smaller the scale at which
it will be effectively dissipated. Since $\phi_{n}$ can grow at most
like $Q^{n\over 3}$ (this corresponds to the extreme case of a fluctuation
carrying a constant energy through the cascade), we conclude that $N$ should
be an integer between $n_{d}$ and $\displaystyle{{4\over 3} n_{d}}$, where we
 have defined $n_{d}$ as
\begin{equation}
n_{d}={3\over 4} \frac{\mbox{Log} R}{\mbox{Log} Q}
\label{defn_d}
\end{equation}
In our numerical study of the statistical properties of the model, we took
$Q=2$, $R=10^{5}$ and let vary $\beta$ between 0 and 1. The choice of 18 shells
 ($N=17$) turned out to ensure the absence of any spurious boundary effect.\\

Since we shall allude sometimes to the complex GOY model, we close this section
by writing down the version of it we used in our computations. With
complex variables $\phi_{n}$ rescaled in the way described just before, the
equations read :
%
%-------------------------- definition des equations du model GOY
\begin{mathletters}
\begin{eqnarray}
\frac{d}{dt}\phi_{0} & = & (2-\epsilon)[1-\phi^{*}_1\phi^{*}_2]-{\phi_0\over
R}\\
\frac{d}{dt}\phi_{1} & = & Q^{2\over
3}[\phi^{*}_0\phi^{*}_2-\phi^{*}_2\phi^{*}_3]
 -{Q^2\over R} \phi_1\\
\frac{d}{dt}\phi_{n} & = & Q^{2n\over
3}[(1-\epsilon)\phi^{*}_{n-1}\phi^{*}_{n-2}
+ \epsilon \phi^{*}_{n-1}\phi^{*}_{n+1} - \phi^{*}_{n+1}\phi^{*}_{n+2}]
 -{Q^{2n}\over R}\phi_{n}\quad 2\leq n \leq N\\
\phi_{N+1} & = & \phi_{N+2}=0
\end{eqnarray}
\label{GOYmod}
\end{mathletters}
%
%
%----------------------------------------------------------------------
Here again the structure of non-linearities is fixed by a single parameter
 $\epsilon$, varying as $\beta$ between 0 and 1. We put for
simplicity the forcing on the $\mbox{zero}^{th}$ shell, as first tried in
\cite{BLLP}. The factor $-i$, which is usually kept in front of the
non-linear terms, as a remnant of the Navier-Stokes equation, has been
absorbed in an innocuous redefinition of variables ($u_{n} \rightarrow
iu_{n}$).
The forcing also can be assumed to be real (here $F_{n}=(2-\epsilon) \delta
_{n,0}$), without loss of generality, thanks to the invariance of equations
under the following phase transformation (see
for instance \cite{BBP})
\begin{mathletters}
\label{GOYsym}
\begin{eqnarray}
\phi_{3n} & \rightarrow & e^{i\theta_{0}} \phi_{3n}\\
\phi_{3n+1} & \rightarrow & e^{i\theta_{1}} \phi_{3n+1}\\
\phi_{3n+2} & \rightarrow & e^{-i(\theta_{0}+\theta_{1})} \phi_{3n+2}
\end{eqnarray}
\end{mathletters}
where $\theta_{0}$ and $\theta_{1}$ are arbitrary angles.
There are obvious differences
 between the real ON model and the complex GOY model, which make their
 comparison interesting in its own right. The phase space which may be explored
by the GOY model is {\em a priori} larger, at least for initial conditions not
 purely real. Also the range of non-linear interactions is wider than in the
first
model. Lastly, the GOY model admits a second quadratic
invariant besides kinetic energy, which is thought to play a prominent role in
fixing its statistical properties \cite{KLWB,GPZ}. Such an extra invariant is
definitely absent in the ON model.\\

\subsection{Fixed points and qualitative description of the phase diagram}

First of all, let us say a few words about the existence and the nature of the
fixed points of the model, which lead to Kolmogorov scaling. In the absence of
dissipation and for an infinite number of shells, $\phi_{n}=1$ is an
obvious solution for any value of $\beta$. However, once viscous dissipation
is introduced, it is easy to see that static solutions making physical sense
can
 exist only for $\beta \neq 0$. Indeed for $\alpha =1$ (OG chain), one has to
solve for every $n>0$
\begin{equation}
Q^{2n\over 3}(\phi_{n-1}\phi_{n} - \phi_{n+1}^2)=\frac{Q^{2n}}{R} \phi_{n}
\label{pfobu_n}
\end{equation}
In order to balance the leading order terms of this equation for $n \rightarrow
+\infty$, $\phi_{n}$ must behave like $-\displaystyle{Q^{4n\over 3}\over R}$
which is not acceptable. This absence of any fixed point in the presence
of dissipation for the Obukhov chain is consistent with the dynamical
behaviour one observes in this case. A solitary wave of negative amplitude
appears invariably at the end of the cascade, which carries back energy
towards large scales. In the final state, energy oscillates back and forth
between the first two shells, the other ones being inactive.\\

Things are different, as soon as a finite amount of Novikov interactions is
 introduced. This is because the term $\phi_{n-1}^{2}$ in equation
(\ref{nlphi_n}) favours energy transfer to small scales. At a static level,
assuming a rapid decay of the spectrum on the ultraviolet side (i.e.
$\phi_{n-1}>>
\phi_{n}>>\phi_{n+1}$), one has now to achieve the balance
\begin{equation}
\beta \phi_{n-1}^{2} \sim {Q^{4n\over 3}\over R}\phi_{n}
\end{equation}
This yields the following general solution
$\phi_{n}=\displaystyle{{Q^{4(n+2)\over 3}\over \beta R}}
\mbox{exp}(-b\,2^{n})$
with $b>0$ otherwise arbitrary. More physically, we may rewrite $\phi_{n}$
in the dissipative range as
\begin{equation}
\phi_{n}={Q^{4(n-n_{d}+2)\over 3}\over \beta}\:\mbox{exp}(-a\,2^{(n-n_{d})})
 \;\;\;\;\;\; \mbox{for} \;n>n_{d}
\label{pfdiss_n}
\end{equation}
with $n_{d}$ given by (\ref{defn_d}). These considerations suggest the
following procedure for computing the fixed point $\phi_{n}^{e}$ in the
presence
of dissipation. The condition of equilibrium for the $n^{th}$ shell being
quadratic in $\phi_{n-1}$, it can be used to express $\phi_{n-1}$ in terms
of $\phi_{n}$ and $\phi_{n+1}$. Going from the last shell with $\phi_{N}$
parameterized as in (\ref{pfdiss_n}) to the first one, and keeping at each step
the positive root, one ends up with $N+1$ numbers $\phi_{N}(a),\ldots,
\phi_{1}(a),\phi_{0}(a)$. The last equation on the $\mbox{zero}^{th}$
shell yields then a solvability condition for the parameter $a$
\begin{equation}
F(a)=1-(\alpha \phi_{1}^{2}(a)+\beta \phi_{0}(a) \phi_{1}(a))=0
\label{pfsolva}
\end{equation}
Figures \ref{shellsta} and \ref{IRbound} show how things work for
$Q=2$, $R=10^{5}$ and $\beta =0.348$. At least in the range of Reynolds numbers
investigated here, there exists only one solution and the resulting fixed point
is drawn on Fig.\ref{ptfix}. The emergence of oscillations of period 2 and
growing amplitude on the infrared side of the spectrum, is a generic feature
of the ON model. Similar phenomena involving period 3 are known to plague the
GOY model. Here the explanation is the following : in the inertial range
(i.e. $1\ll n\ll n_{d}$), the equation to be solved reduces to
\begin{equation}
q_n^{-1}(1+Cq_n^{-1})-q_{n+1}(q_{n+1}+C)=0
\label{itermap}
\end{equation}
where one has introduced the ratio $q_n={\phi_n\over \phi_{n-1}}$ and $C={\beta
\over \alpha}$. Equation
(\ref{itermap}) defines a map $q_{n}=g(q_{n+1})$, whose fixed point $q_{n}=1$
is easily seen to be unstable for $C\leq 1$. Only for discrete values of the
Reynolds number (such that one of the crossing points in Fig.\ref{shellsta}
has ordinate 1), can one hope to get rid of these oscillations. We should say
that this odd-even disymmetry, though a pathology of the ON model, is less
visible
in the chaotic state to be described below. Furthermore, it does not affect
other
physical quantities like the energy transfer $\epsilon_{n}$
defined in (\ref{flux_n}).\\

We turn now to a qualitative discussion of the phase diagram of the model,
which is observed as $\beta$ varies (for the particular values $Q=2$ and
$R=10^{5}$). Although the static solution exists for every strictly positive
 $\beta$, it becomes unstable for $\beta \leq 0.355 \pm 10^{-3}$ and evolves
 towards a periodic limit cycle through a first Hopf bifurcation. A scenario
\`a la Ruelle Takens, similar to the one already discovered in the GOY model
 \cite{BLLP}, leads then to chaos for $\beta \leq \beta^{*} = 0.349 \pm
 10^{-3}$. We did not try to get very precise estimates of these two
thresholds,
 whose position
is expected to vary with $R$. From a slightly unrelated analysis of
 movable singularities of the ON model, presented in Section 5, we shall
 speculate later in the paper that the asymptotic value of $\beta^{*}$ in the
 limit of infinite Reynolds number is of order 0.394. In the quasiperiodic
 regime,
the shells oscillate in a coherent way around the static fixed point
 discussed above. Their oscillations remain of moderate amplitude even close to
the transition, and one has with a very good accuracy, for all integers $n$ and
 $p$, $\langle \phi_{n}^{p}(t)\rangle =(\phi_{n}^{e})^{p}$
(where $\langle  \ldots \rangle$ denotes a temporal average).\\
The behaviour in the chaotic phase is rather simple to understand, far from
the transition. For values of $\beta$ not greater than say 0.28, the dynamics
consists in well isolated pulses emitted from the forced shell, after it has
reached a significant level. The pulse propagates down the scales, through
almost inactive shells, leaving behind a finite amount of energy. After being
stopped by dissipation, it gives rise to a rather well characterized pulse
of negative amplitude, which carries back to the large scales most of the
energy
that dissipation failed to absorb. This is quite close to what is observed
in the Obukhov limit, although here things repeat themselves in a slightly
 disordered manner. The picture gets more complicated as $\beta$ gets closer
to $\beta^{*}$ (typically for $0.30\leq \beta \leq 0.349$). Shells in the
inertial
range remain almost always active and they form a noisy background
out of which only the biggest fluctuations develop like singular pulses.
 Negative excursions of variables, triggering backflow of energy, occur very
rarely. Splitting of pulses as well as partial reflection of some of them may
be observed all the way along the cascade. The next Section will help to
 quantify a bit these statements.\\
\section{Multiscaling properties of the chaotic phase}

Moments of the variables $\phi_{n}$, defined as
 $S_{n,p}=\langle\phi_{n}^{p}\rangle$,
are good tools for characterizing the intermittency of shell models. According
to (\ref{defphi}), exponential growth
of any of the $S_{n,p}$'s with $n$ in the inertial range is the sign of
 deviations from K41-scaling. However we shall rather concentrate on moments of
the energy flux $\epsilon_{n}$ defined in (\ref{flux_n}), namely on quantities
$\Sigma _{n,p}=\langle |\epsilon _{n}|^{p\over 3}\rangle$
The reason for this choice has already been discussed in \cite{KLWB}
within the context of the GOY model.
The point is that for moderate $p$, the oscillations we found in the static
 solution still contaminate $S_{n,p}$. Although they are much smaller, they
 prevent us from an accurate determination of the scaling exponents. As far as
scaling properties are concerned, the $\Sigma_{n,p}$'s provide a valuable
 alternative, because they are free from any spurious oscillations.\\
On the other hand one must be aware that the absolute value introduces a
 considerable bias for the lowest moments if the energy cascade has a poor
positive "smooth component",  as it is the case far away from the transition
 where the mean energy transfer is very small. When on the contrary the smooth
 component is important, $\epsilon _n$ is almost always positive and
$\langle |\epsilon_{n}|\rangle$ is very close to a constant in the inertial
range as $\langle \epsilon_{n}\rangle $ should be in a statistically
stationnary
 state. In any case the bias disappears at higher values of $p$ since the
 strongest fluctuations are always positive.\\
 We have calculated $\Sigma_{n,p}$ for $p$ up to $12$. For each run we
 integrated the equations over $1000$ turn-over or unit times and sampled the
 signal with a step much smaller than the characteristic time-scales of the
more intense structures (in practice we took $\Delta t =10^{-4})$. The length
of
 each run was enough to ensure the stationarity
 of the statistics, at least for moments of order $p$ smaller than $7-8$.
As shown on Fig. \ref{exscal} for $\beta =0.33$, these quantities obey nice
 power laws on a rather wide range of shell numbers. We may thus define
 exponents  $\sigma_{p}$ such that in the inertial range
$\Sigma_{n,p}\sim {k_{n}}^{\sigma_{p}}$
(the relation between $\sigma_{p}$ and the usual exponent $\zeta_{p}$
associated to the velocity field is : $\zeta_{p}={p\over 3}-\sigma_{p}$).\\

To extract values of the scaling exponents  $\sigma_{p}$ from our data,
we again followed the procedure outlined in \cite{KLWB} (though not with the
 same refinement!).
The inertial or fitting range was determined as the
interval of values of $n$, for which a least square fit of the data to a
straight line, give $\sigma_{1}$ the closest to zero. A
fitting range $5\leq n \leq 10$ proved to be the best for all values of
$\beta$ we looked at, with $\sigma_{1}$ as small as $0.01$ for $\beta \sim
 0.33$. Actually the highest moments ($p\geq7$) allow a wider range for an
 estimation of the exponents, whereas the lowest depend only slightly on this
 length. In order to get a rough estimate of statistical errors due to the
 finite length of our temporal signal, we repeated the same operations many
 times (typically $5$), taking as new initial conditions the state obtained at
 the end of the precedent run. Error bars drawn in the next figures take into
account this only source of uncertainty. Our results for the scaling
exponents  $\sigma_{p}$ are summarized in Fig.\ \ref{tp1} for three
typical values of $\beta $ : $\beta =0.28$, $\beta =0.33$, and $\beta
=0.343$.\\
As concerning the scattering of data, we see that it remains small for
$p\leq 8$, except when we get very close to the transition, a not too
surprising
fact. The overall shape of the  $\sigma_{p}$-curve illustrates the
distinction made at the end of Section 2 between two chaotic regimes. For
$\beta =0.28$, i.e. rather far from the transition, one crosses over rapidly
($p\geq 5$) towards a linear growth of exponents with $p$. The exponents
take rather big negative values for $p\leq 3$, presumably because of the
important role played by energy backflows in this case. For the two other
values of $\beta$ we investigated, the cross-over region is significantly
wider.
There is clearly curvature and henceforth multifractality, even if for
$p \geq 7-8$ an asymptotic linear regime ultimately sets in.
Figure \ref{tp2} offers a magnified view on the cross-over region. The fact
that exponents still vary strongly between $\beta =0.33$ and $\beta =0.343$
 proves
the influence of the proximity of the transition on the physics probed
by the moments of corresponding order. On the other hand, it can be noted that
the asymptotic slope remains almost the same. This feature will be explained
in the next Section in terms of self-similar solutions parameterizing the
biggest fluctuations of our hydrodynamic system.\\

As a way of checking our numerics, and also in order to get some insight
into the differences
 between the two models, we have performed strictly analogous computations on
 the complex GOY model for two different values of $\epsilon$ (with $Q=2$,
$R=10^{5}$ and $N=17$). The first one,
$\epsilon =0.75$, lies far from the transition, which was found in \cite{BLLP}
to occur at $\epsilon^{*}=0.395..$ for a Reynolds number roughly
the halfth of ours. The second one is the standard $\epsilon=0.5$, known to
lead to scaling properties in good agreement with experiment \cite{JPV}.
Figure \ref{GOYON1} shows on the same graph the scaling exponents for the ON
 model
with $\beta =0.28$ and the GOY one for $\epsilon =0.75$. The resemblance
is striking, apart from from slightly different asymptotic slopes of the two
 curves and a cross-over region a bit wider for the GOY model. In Fig.\
 \ref{GOYON2}, the same comparison is made between the two models, this
 time for $\beta =0.33$ and $\epsilon =0.5$. Now differences show up,
especially at large orders. While a linear monofractal behaviour has
definitely set in for $p\geq 7$ in the case of the ON model, the local slope
 of the $\sigma_{p}$ does not stop increasing in the GOY model, even though
 this is hardly perceptible to the eye. Also significant is the lack of
convergence for $p\geq 8$. We note that the points marking the upper
error bars, which were obtained from one particular run among six of
equal temporal length, are surprisingly well fitted by the formula
proposed by She and Leveque \cite{SL} for real Navier-Stokes turbulence
\begin{equation}
\sigma_{p}=\frac{2p}{9} -2[1-(\frac{2}{3})^{p\over 3}]
\label{SheLev}
\end{equation}
up to the highest values of $p$ investigated here. A physical reason for this
good agreement observed by many others is missing.
We did not try to tighten our error bars by
 increasing the length of numerical integration since this was beyond the scope
 of this work. The results of the next Section will somehow confirm the
peculiar nature of intermittency in the GOY model for such values of
parameters as $\epsilon =0.5$ and $Q=2$.\\

\section{Study of self-similar solutions}

Let us take for granted from the results of the preceding Section that the
scaling exponents $\sigma_{p}$ grow asymptotically like $\gamma p$
at large $p$, with $\gamma $ a positive number depending on the parameter
$\beta $ only. This means that the amplitude of fluctuations carrying the
system away from the K41 fixed point cannot grow from shell to shell more
rapidly than $Q^{\gamma n}$. On the other hand, the fact that $\gamma$ takes
a finite value even close to the transition suggest that such fluctuations
are efficient as soon as the instability threshold is passed. It is the
purpose of this Section to identify the set of singular fluctuations that
the ON model can admit.\\
Since we are now interested in nonlinear instabilities
occuring in the inertial range, we may forget about forcing and dissipation,
and think of the shell number $n$ as running from $-\infty$ to $+\infty$.
Let us rewrite the equation of motion in terms of new variables
$b_{n}=Q^{n}u_{n}=Q^{2n\over 3}\phi_{n}$ ($b_{n}$ is nothing dimensionally
but the gradient of the velocity field). We get from (\ref{nlphi_n})
and (\ref{equphi_n}), after absorbing the factor $\alpha Q^{2\over 3}$ into
a rescaling of time :
\begin{equation}
\frac{d}{dt} b_{n}= N_{n}[\vec{b}]=(b_{n-1}b_{n}+{\beta\over \alpha}
Q^{2\over 3} b_{n-1}^2) - {1\over Q^{2}}(b_{n+1}^2+{\beta\over \alpha}
Q^{2\over 3}b_{n}b_{n+1})
\label{nlb_n}
\end{equation}
Since ${\vec N}[\vec b]$ does not depend explicitely on $n$ and is
quadratic in $\vec b$, the set of equations (\ref{nlb_n}), for $-\infty
< n< +\infty$, support formally self-similar solutions of the type :
\begin{equation}
b_{n}(t)={1\over {t^{*}-t}} f(Q^{nz}(t^{*}-t))\;\equiv
Q^{nz} g(Q^{nz}(t^{*}-t))
\label{selfsim_1}
\end{equation}
In the equation above, $t^{*}$ is the critical time at which, in the absence
of dissipation, the fluctuation reaches the end of the cascade. The scaling
exponent $z$ is {\em a priori} arbitrary. However $z={2\over 3}$ gives
back Kolmogorov scaling, while $z=1$ corresponds to the extreme situation
of a fluctuation carrying a constant energy. One expects therefore
${2\over 3}\leq z \leq 1$ on physical grounds.\\
Self-similar solutions, if they exist, are obviously good candidates for
describing the growth of singular fluctuations. The question then arises
whether many values of $z$ are dynamically accessible, which would be a natural
source of multifractality, or whether on the contrary a single $z$ is selected.
In that case, one should check that $\gamma =(z-{2\over 3})$, since moments
$\Sigma_{n,p}$ are dominated by extreme fluctuations for high values
of $p$. Self-similar solutions together with their exponent $z$ have
already been determined by Nakano for the ON model \cite{N88}. He used a rather
cumbersome iterative method to find them and we were not convinced he
had exhausted the whole set of possibilities in his work. This is why
we came back to this problem and were led to develop a procedure to be
described below, which is quite efficient and easily extended to any shell
 model. It should however be said from the beginning that our results about the
ON model are in complete agreement with the conclusions reached in
\cite{N88}.\\

By plugging the Ansatz (\ref{selfsim_1}) into (\ref{nlb_n}) and introducing
 the logarithmic variable $\xi =n+{1\over z \log Q}\log (t^{*}-t)$, one arrives
at the following equation for $f$
($f$ is actually divided by $z \log Q$ to make the result a bit simpler)
\begin{equation}
f'(\xi )-z\log Q f(\xi )=(f(\xi -1)f(\xi )+{\beta\over \alpha}
Q^{2\over 3} f(\xi-1)^{2}) - {1\over Q^{2}}(f(\xi+1)^{2}+{\beta\over \alpha}
Q^{2\over 3}f(\xi )f(\xi+1))
\label{selfsimequ}
\end{equation}
If square integrability of $f$ is required, Eq.\ (\ref{selfsimequ}) is
nothing but a non-linear eigenvalue problem for the unknown $z$,
which is very difficult to solve directly, either
analytically or numerically. To make progress, we can try to approach
$f$ dynamically. Rather than coming back to the original equations
of the model, let us introduce a fictitious dynamics leaving the norm of
the $(N+1)$-dimensional vector $\vec b$ invariant
\begin{equation}
\frac{d}{d\tau} \vec{b}= \vec{N}[\vec{b}]-\frac{<\vec{N}[\vec{b}],\vec{b}>}
{<\vec{b},\vec{b}>}\, \vec{b}
\label{projdyn}
\end{equation}
In the equation above,
$\vec{N}[\vec{b}]$ is the vector of components $N_{n}[\vec{b}]$, whose
expression was given in Eq. (\ref{nlb_n})
%(with the convention $b_{-1}=b_{N+1}=0$)
and $\langle \vec{A},\vec{B}\rangle =\sum_{n=0}^{N} A_{n}B_{n}$ is
the usual euclidean scalar product. The projection factor, which intervenes in
 the r.h.s. of (\ref{projdyn}) to keep $\vec b$ on a sphere~:
\begin{equation}
A(\tau)=\frac{<\vec{N}[\vec{b}](\tau),\vec{b}(\tau)>}
{<\vec{b}(\tau),\vec{b}(\tau)>}
\label{defA}
\end{equation}
will be of central importance in the following.\\
Characteristic time scales on shell $n$ are in first approximation proportional
 to $b_{n}$ but now $b_{n}$ cannot exceed the initial value of
$\sqrt{<\vec{b},\vec{b}>}$. It follows that within the "projected dynamics"
defined by Eq.\ (\ref{projdyn}), the cascade towards small scales is not
accompanied by an acceleration of motion as in the original equations.
There is now no impediment
against taking a very large number of shells since the required time
resolution does not grow anymore exponentially with $N$. By integrating
numerically (\ref{projdyn}), we observed that any
initial condition of finite support (i.e. $b_{n}(0)\neq 0$ for $0\leq n \leq
n_{0}$, with $n_{0}\ll N$) gives birth at large times $\tau$ to a solitary
 wave moving with a constant velocity towards small scales.
In other words, a period $T$ may be defined such that asymptotically, for $\tau
\rightarrow +\infty$ (a more precise condition reads $1\ll \tau \ll NT$,
because some reflection will ultimately occur on the ultraviolet boundary),
\begin{equation}
b_{n+1}(\tau +T) =  b_{n}(\tau) \;\;\equiv \;\;
b_{n}(\tau)=b(n-\frac{\tau}{T})
\label{asymptb_n}
\end{equation}
Note that (\ref{asymptb_n}) implies $A(\tau +T)= A(\tau)$.
The shape of the final solitary wave is found to be always the same, up to
the scaling symmetry
\begin{equation}
b(n-{\tau \over T}) \rightarrow \lambda b(n-\lambda{\tau \over T})
\label{scalesym}
\end{equation}
and it is remarkably stable, as demonstrated
by Figs.\ \ref{selfsim1} and \ref{selfsim2}.\\
Let us now make the connection between this finding and self-similar solutions
in shell models. This is easily done by writing any solution $\vec{b}(\tau)$ of
Eq. (\ref{projdyn}) in the form
\begin{equation}
\vec{b}(\tau)=\exp \bigl(-\int_{0}^{\tau} \! A(\tau ') d\tau '\bigr)\,
 \vec{c}(\tau)=B(\tau) \vec{c}(\tau)
\label{transfproj}
\end{equation}
Since the non-linear kernel is quadratic, one gets for $\vec{c}(\tau)$
\[\frac{d}{d\tau}\vec{c}=B(\tau) \vec{N}[\vec{c}\,]\]
The original dynamics : $\displaystyle{\frac{d}{dt}\vec{c}}=\vec{N}[\vec{c}\,]$
is recovered, after defining the physical time $t$ as
\begin{equation}
t(\tau)=\int_{0}^{\tau} \! B(\tau ') d\tau '
\label{transftime}
\end {equation}
These straightforward manipulations prove that every solution $\vec{b}(\tau)$
of Eq. (\ref{projdyn}) can be mapped onto a solution $\vec{c}(t)$ of the
real physical problem in the inertial range, according to the transformation
law~:
\begin{equation}
\vec{c}(t)=\exp \bigl(\int_{0}^{\tau (t)} \! A(\tau ') d\tau '\bigr)\,
 \vec{b}(\tau (t))
\label{transfsol}
\end{equation}
where $\tau (t)$ is obtained from the inversion of Eq. (\ref{transftime}). It
can now be seen that a travelling wave in the projected
 dynamics, of period $T$ and average value in time $\langle A(\tau)\rangle >0$,
 is the signature of a self-similar solution in the true dynamics. Indeed,
 according to (\ref{transfsol}), each time the component of $\vec{b}$ of
maximal
 amplitude moves from one shell to the next, $\vec{c}$ is multiplied by
$\exp (\langle A\rangle T)$. From a comparison with the initial Ansatz
 (\ref{selfsim_1}), one gets :
\[ Q^{z} = \exp (\langle A\rangle T) \]
or
\begin{equation}
z=\frac{\langle A\rangle T}{\log Q}
\label{computz}
\end{equation}
This formula allows one to obtain accurate estimates for $z$, since both
quantities $\langle A\rangle $ and $T$ are easily measurable (and their product
is left invariant as it should by the scaling symmetry (\ref{scalesym})). \\

The method was first used to compute $z$ for various values of $\beta$ in the
ON model. Results are summarized in table \ref{tab:1}, where a comparison
between $z-{2\over 3}$ and the asymptotic slope $\gamma$ of the $\sigma_{p}$-
curve is also made. We find a reasonable agreement between these last two
 quantities, in view of the comparatively large errors in the estimate for
$\gamma$. It is important to
realize that the existence of self-similar solutions has nothing to do
with the presence of chaos. In the ON model, the exponent $z$ decreases
gently from 1 to $\displaystyle{2\over 3}$, as $\beta$ varies between 0.145..
and 1 (as already noticed in \cite{S78,N88}, one has $z=1$ for $\beta \leq
 0.145..$). The analytic stucture of the solitary wave remains the same. The
scaling function $f$ of Eq. (\ref{selfsimequ}) presents an essential
singularity
 $f(\xi) \sim 2^{\xi} \exp (-2^{\xi})$ for $\xi \rightarrow +\infty$
(up to subdominant multiplicative corrections), and an exponential tail
 $f(\xi) \sim Q^{z\xi}$ for $\xi \rightarrow -\infty$.
In the case where $z=1$, the exponential tail is replaced by a second essential
singularity $f(\xi) \sim 2^{-\xi} \exp (-2^{-\xi})$. Table
\ref{tab:1} shows that $z \sim 0.88$ at the transition between the regular
and chaotic regimes, located near
$\beta =0.349$. This high value explains why the ON model (at least for
$Q=2$) is bound to exhibit rather strong intermittency in the chaotic part
of its phase diagram.\\
We were curious to extend this analysis to the complex GOY model. It is a
simple matter to generalize Eq.\ (\ref{projdyn}) to the case of a complex
vector. Details will not be given here. The conclusion of our (partial)
investigations is that the GOY
model also possesses only one ideal self-similar solution for a given value
of $\epsilon$. Furthermore, this self-similar solution is purely real and
positive, up to the phase symmetry (\ref{GOYsym}) of the model. This means
 that the complex amplitudes $b_{n}(\tau)$ take the
 asymptotic form  $b_{n}(\tau)=e^{i\theta (n)}b(n-{\tau \over T})$, where
 the phase $\theta (n)$, subject to the constraint
 $\theta(n)+\theta(n+1)+\theta(n+2)=0$, is the only footprint of the initial
condition and the amplitude $b$ presents a shape quite similar to the one
 obtained for the ON model.
 Quantitative results are presented in the table \ref{tab:2}. The
comparison between large order statistics and scaling properties of
self-similar
solutions was done only for two values of $\epsilon$~: $\epsilon=0.75$ and
$\epsilon =0.5$. While in the former case the same agreement is obtained as for
the ON model, we find in the latter a discrepancy by a factor 2 between
$\gamma $ and $z-{2\over 3}$. The discrepancy is even bigger if one
extrapolates from the She-Leveque formula (\ref{SheLev}) $\gamma =2/9=0.222$.
We think that the failure of self-similar solutions to explain intermittency
at high orders in this case, lies in the closeness to the Kolmogorov value
$2\over 3$ of their scaling exponent $z$. After all, with $z-{2\over 3}$ as
small as 0.052 and a Reynolds number $R=10^{5}$ as in our computations, the
amplitude of singular fluctuations grows, upon propagating from the integral
scale to the dissipative one, by a factor $Q^{(z-{2\over 3})n_{d}}
=R^{{3\over 4}(z-{2\over 3})}$, which does not exceed 1.5 ! This gives very
little chance to such a fluctuation to survive collisions with the turbulent
background and to govern statistics at large orders. We find it plausible
that the mildness of singular fluctuations and the finite length of the cascade
combine to produce a new kind of intermittency with a more pronounced
multifractal character. It is an interesting issue, left for further
investigation, to understand how the system is then able to develop
an asymptotic growth of the $\sigma_{p}$ with $p$ steeper than the one
expected on the basis of self-similar solutions.\\

\section{Movable singularities as a signature of chaos}

{}From a formal point of view, self-similar solutions studied in the
previous Section describe the approach of the system towards blowing-up,
which, in the absence of dissipation, happens in finite time. It is also of
interest in the context of nonlinear o.d.e.'s to consider movable
singularities taking place at complex times. The local structure of such
objects is intimately linked to the non-linearity, while their distribution
in the complex $t$-plane may help to understand such physical properties as
high-frequency intermittency \cite{FM}. Besides, according to Painlev\'e's
criterion, non algebraic singularities indicate usually lack of
 integrability.
This yields a very economical way to detect analytically the presence
of chaos in any dynamical system (see for instance \cite{DFHGMS}
and the references therein related to this topic). There are two main
reasons why we report in this Section a study of movable singularities in
the ON model, which at first sight is disconnected from the rest of the paper.
 The first one is purely technical : it turns out that the
method used to determine the local structure of movable singularities
(and possibly their position) is quite close to the one developed in Section
 4 for tracking self-similar solutions. The second reason has more to do with
physics~: whereas self-similar solutions by themselves had nothing to tell us
 about the chaotic properties of the model, we shall see that movable
singularities in the complex $t$-plane disappear (or better said, get trapped
on the last shells near the ultraviolet boundary) as $\beta$ exceeds a
value of order $0.394\pm 10^{-3}$. It is tempting  to speculate that this
threshold marks the ultimate boundary between chaotic and
regular dynamics, the one reached in the limit of infinite Reynolds number.
We shall also get strong indications that movable singularities in shell
models parameterize energy backflows and as such could be responsible
for the peeling off of coherent structures as they cascade downwards
to small scales. It will become rapidly clear to the reader that the analysis
to be presented below, though restricted to the ON model, can easily
be applied to any shell model with presumably similar conclusions at the end.\\

We shall work with the vector $\vec b$ defined in Section 4. The
quadratic degree of non-linearities implies that the only movable
singularities are poles so that~:
\begin{equation}
b_{n}(t) \sim \frac{a_{n}}{t-t^{*}} \;\;\; \mbox{for}\;\;\;
 t\rightarrow t^{*}\;\;\;\mbox{and}\;\;\; 0\leq n\leq N
\label{defpole}
\end{equation}
where $t^{*}$ is an arbitrary complex critical time. The $N+1$ residues
$a_{n}$ form a vector $\vec a$, which after substituting (\ref{defpole}) into
(\ref{projdyn}) is seen to obey the condition~:
\begin{equation}
- \vec{a}=\vec{N}[\vec{a}\,]
\label{equapole}
\end{equation}
The problem now is to solve (\ref{equapole}). This is a much more difficult
task than computing fixed points as in Section 2. First, $\vec a$ is
necessarily complex (it is easy to check that (\ref{equapole}) implies
$\sum_{n=0}^{N} a_{n}^{2} Q^{-2n} =0$). Second, we expect on physical
grounds the vector $\vec a$ to be localized in shell space. This means that
we are looking for solutions of Eq. (\ref{equapole}) which would be
square-summable ($\sum_{-\infty}^{+\infty} |a_{n}|^{2} < +\infty $), were the
range of shell numbers extended to the whole set of relative integers.
Any "shooting" method of the type outlined in Section 2,
which would start from one endpoint and try to join the other one with the
appropriate asymptotic behaviour, is in fact doomed to failure because of
strong numerical instabilities.\\
As in the preceding Section the idea will be to approach dynamically the
desired solutions to (\ref{equapole}). Before doing so, we must say a few words
about the notion of "genericity" of movable singularities. Consider a
 singularity at time $t^{*}$ and assume $\vec a$ is known. Equation
 (\ref{defpole}) gives only the leading order term in the expansion of $\vec b$
near $t^{*}$, which may be pursued order by order just from local analysis.
Writing $\vec b$ as $\displaystyle{{\vec{a}\over t-t^{*}} + \delta \vec{b}}$,
where the correction $\delta \vec{b}$ is small compared to the
$\mbox{zero}^{th}$ order term, one gets to linear order in $\delta \vec{b}$
\begin{equation}
\frac{d}{dt} \delta \vec{b}=\frac{1}{t-t^{*}} M \delta \vec{b} +\vec{F}
-\frac{\vec{D}[\vec{a}]}{t-t^{*}}
\label{equacorr}
\end{equation}
In the equation above, $M=\left[\displaystyle{{\partial N_{i}\over \partial
 b_{j}}}\right]$ is the Jacobian matrix of the nonlinear kernel
$\vec{N}[\vec{b}]$ evaluated at point $\vec a$.
Forcing and dissipation were kept for completeness in the right hand side of
 (\ref{equacorr}) but only the homogeneous part of the equation really matters
in what follows. It has (N+1) independent solutions of
the form $(t-t^{*})^{\mu_{i}} \vec{b}_{i}$, where $\mu_{i}$ is the $i^{th}$
eigenvalue of $M$ and $\vec{b}_{i}$ the corresponding eigenvector.
Provided $Re\, \mu_{i} >-1$, a correction of the type $\lambda_{i}
(t-t^{*})^{\mu_{i}} \vec{b}_{i}$ with $\lambda_{i}$ an arbitrary complex
number, is free to appear in the expansion of $\vec b$ around
$t^{*}$, since it is indeed smaller than the $\mbox{zero}^{th}$ term.
Actually, $\vec N$ being quadratic, $\vec{N}[\vec{a}]=-\vec{a}$ implies
$M\vec{a}=-2\vec{a}$. Therefore, one of the $\mu_{i}$ (say $\mu_{0}$) equals
by construction -2. The eigenvalue -2 and the corresponding eigendirection
$\vec{a}$ are associated to the arbitrary position of $t^{*}$ and as such
must be excluded from the expansion of $\vec{b}$. It follows that the
most general expression of $\vec{b}$ around a singularity reads~:
\begin{equation}
\vec{b}(t)=\frac{\vec{a}}{t-t^{*}} +\sum_{i=1}^{N_{s}} \lambda_{i}
(t-t^{*})^{\mu_{i}} \vec{b}_{i} +\mbox{h.o.t.}
\label{genexp}
\end{equation}
where $N_{s}$ is the number of eigenvalues of $M$, whose real part is bigger
than -1. It is not difficult to check that, once the $N_{s}$ complex numbers
$\lambda_{i}$ are given, there is no arbitrariness left in the rest of the
expansion (denoted as h.o.t. in (\ref{genexp})). What we have in our hands is
a local
expression of our solution which depends on $(N_{s}+1)$ parameters ($t^{*},
\lambda_{1},\ldots \lambda_{N_{s}}$), whereas $(N+1)$ initial conditions
are necessary to specify entirely the evolution of the dymamical system.
Therefore a singularity will be generic (i.e.\ it will not result from a set
of initial conditions of zero measure), if and only if $N_{s}=N$. In other
words, we are interested only in solutions to (\ref{equapole}) with
$N$ eigenvalues $\mu_{i}$ of real part bigger than -1, besides the
trivial one $\mu_{0}=-2$. \\

The previous considerations suggest the introduction of the following
dynamics~:
\begin{equation}
\frac{d}{d\tau}\vec{a}=-\vec{N}[\vec{a}]+\bigl(Re \frac{\langle \vec{a},
\vec{N}[\vec{a}]\rangle }{\langle \vec{a},\vec{a}\rangle } + i\delta\,Im
\frac{\langle \vec{a},\vec{N}[\vec{a}]\rangle}{\langle \vec{a},\vec{a}\rangle}
\bigr) \vec{a}
\label{residyn}
\end{equation}
where, since we are dealing now with complex-valued vectors,
$\langle \vec{A},\vec{B}\rangle =\sum_{n=0}^{N} A^{*}_{n}B_{n}$
The second term in the r.h.s.\ of (\ref{residyn}) keeps the norm of $\vec{a}$
constant. The last one affects only its phase and one
is in principle free to choose any value for the parameter $\delta$.
It may be shown that there is a one-to-one correspondance between
fixed points of the dynamics (\ref{residyn}) with a basin of attraction
of finite measure, and generic solutions (in the sense of the previous
paragraph) to the initial problem. A proof of this almost intuitive statement
is given in the Appendix. It has nice consequences : in order to determine
the possible arrangements of residues $a_{n}$, it suffices to integrate
 (\ref{residyn}) for initial conditions which are not purely real (otherwise
 they remain so forever). If after a long enough time, a stationary state
 $\vec{a}_{f}$ is reached, then~:
\begin{equation}
\vec{a}=-\vec{a}_{f} \displaystyle{\frac{\langle
 \vec{a}_{f},\vec{a}_{f} \rangle}{\langle
 \vec{N}[\vec{a}_{f}],\vec{a}_{f}\rangle}}
\label{resirescal}
\end{equation}
contains the desired information.
Note that the computational cost of the method increases only linearly
with the number of shells $N+1$. It is therefore easy to get rid
of finite size effects if necessary.\\
We have applied this technique to the ON model and made the following
observations. As anticipated on the basis of the preceding considerations,
 the vector $\vec{a}$ evolves systematically towards a fixed point
provided the condition $\delta \geq 1$ is met (actually, the marginal case
$\delta =1$ still works but requires longer times of integration). After
performing the rescaling (\ref{resirescal}), the final state of $\vec{a}$
(giving access to the residues $a_{n}$) was found to be always the same, up to
complex conjugation (which is an obvious symmetry of (\ref{equapole}))
 and translation along the shell number axis.
This last property, which is crucial to ensure the "mobility" of the
 singularity in momentum space, holds for $\beta < \beta^{*}=0.394\,\pm
 10^{-3}$. For $\beta \geq \beta^{*}$ we find only one  solution, rigidly
 attached to the last shell. Figures \ref{resi_1} and \ref{resi_2} summarize
 the phenomenon by showing the modulus and the real
 part of $a_{n}$ for respectively $\beta =0.39$ and $\beta =0.40$. They
 were deduced from a numerical integration of (\ref{residyn}) with $N=29$,
 $\delta =2$ and the initial condition $a_{n}=i \delta_{n,0}$. The imaginary
 part of $a_{n}$ has not been represented in order not to burden the
 figures. For $\beta =0.39$, a change in the initial conditions or in the value
 of $\delta$ most likely leads to a displacement of the peak of the final
 structure along the horizontal axis. In contrast, for $\beta =0.40$, the peak
resides always on the last shell. A perfect convergence onto a true fixed
point of (\ref{residyn}) is
difficult to achieve because of slow transients near the transition. Thus we
cannot exclude some minor adjustments of residues with respect to the
picture shown here, especially at the rear end of the structure ($n\geq 10$
in Fig.\ \ref{resi_1}). Note the characteristic pattern at the front ($5\leq n
\leq 7$ in Fig.\ \ref{resi_1}) with a large negative excursion of $Re(a_{n})$,
which by the way may be still recognized in Fig.\ \ref{resi_2}, i.\ e.\
 beyond the threshold. From a mathematical point of view, solutions in the
inertial range, as depicted by Fig.\ \ref{resi_1}, disappear when one of the
 eigenvalues $\mu_{i}$ of the Jacobian matrix $M$ gets a real part smaller
 than -1. Apparently, the only place where they manage to survive is near
 the ultraviolet boundary, where the nonlinear kernel is strongly
 modified.\\
We shall not expand too much on these findings. At least they prove
that complex time singularities are not involved in the building-up
of self-similar solutions, because in contrast to the former, the latter were
found to exist for any value of $\beta$. Just from this obvious remark, it is
 tempting to infer that complex time singularities in shell models, when
sufficiently close to the real time axis, encode the occurence of "blockades"
in the energy cascade, leading possibly to negative excursions of shell
 amplitudes and more or less developed energy backflows. This interpretation is
corroborated by the wild oscillations displayed by the phase of residues
and also the fact that such objects form most naturally at the ultraviolet
boundary as suggested by Fig.\ \ref{resi_2}. The system is bound to exhibit
regular dynamics for $\beta \geq \beta ^{*}$ because it has lost these
 agents of disorder.\\

Before closing this Section, we would like to mention that equations (\ref
{residyn}), which were introduced as an abstract auxiliary tool,  may also
 be used more concretely for locating singularities of a real
 solution of the shell model (neglecting forcing and dissipation).
Consider indeed initial conditions of the form $\vec{a}=i\vec{b}_{0}$
where $\vec{b}_{0}$ is arbitrary but real. It may be checked, by using
manipulations similar to those leading in Section 4 to Eq.\ (\ref{transfsol}),
that stepping forward the fictitious dynamics (\ref{residyn}) is in fact
 equivalent to integrating the original dynamics
 $\displaystyle{\frac{d}{dt}}\vec{b}=\vec{N}[\vec{b}]$
(from the initial condition $\vec{b}_{0}$) along a trajectory in time space
 parameterized as~:
\begin{equation}
t=-i \int_{0}^{\tau }\! \mbox{exp}\bigl(\int_{0}^{\tau '} A(\tau '')d\tau ''
\bigr) d\tau '
\label{cmplxtrajec}
\end{equation}
where $A(\tau )$ reads~:
\begin{equation}
A(\tau )=Re \frac{\langle \vec{a}(\tau ),\vec{N}[\vec{a}](\tau )\rangle }
{\langle \vec{a}(\tau ),\vec{a}(\tau )\rangle } + i\delta\,Im
\frac{\langle \vec{a}(\tau ),\vec{N}[\vec{a}](\tau )\rangle}
{\langle \vec{a}(\tau ),\vec{a}(\tau )\rangle}
\end{equation}
For $\delta =0$, $A(\tau )$ is real and according to
(\ref{cmplxtrajec}) the path followed in the complex time plane
is parallel to the imaginary axis. The probability of crossing a singularity
in this way is obviously null for arbitrary initial conditions. For finite
values of $\delta $, the trajectory gets curved in such a way that,
 for $\delta > 1$, it finds with probability one a singularity of $\vec{b}(t)$
 at the end.\\

\section{\bf Conclusion}

Starting from a numerical investigation of the ON model, we were led
to identify elementary bricks in its dynamics, which must exist more generally
 in any scalar shell model. Interestingly enough, they appear to have
rather constrained structures. Naturally the construction of a statistical
theory from these deterministic objects remains a hard task. But we think that
a
 precise knowledge of their properties may help to formulate new questions.
For instance the discrepancy found in the case of the GOY model for
$\epsilon =0.5$ between the asymptotic growth of scaling exponents of
statistical moments and the strength of extreme fluctuations is a puzzling
fact, which clearly deserves further investigation. Another isssue concerns
the selection mechanism of the scaling exponent
$z$ of self-similar solutions whose present understanding is still poor. One
 must remember that the method developed in the paper is in essence dynamical.
 We cannot therefore exclude the existence of a larger manifold of solutions,
out of which only the element with the smallest $z$ would be systematically
observed. Clearly more mathematically oriented work would be welcome to
elucidate this technical point, which may be of some physical relevance.\\
%We plan to explore in the next future the complete phase diagram of the ON
%model in the ($\beta ,Q)$ plane, with the hope of finding chaotic regions
%associated to an exponent $z$ close to $2/3$. This would allow a better
%%%assessment of the respective properties of the real ON and complex GOY
%models.\\

\noindent
{\bf Acknowledgment}\\
We are grateful to E. Gledzer for suggesting us to revisit the ON model.
He was a constant source of inspiration during the course of this work, which
benefitted also from many discussions with B. Castaing and Y. Gagne.\\

\begin{figure}
\caption{The amplitudes of the first two shells as functions of the parameter
$a$ describing the dissipative range.}
\label{shellsta}
\end{figure}

\begin{figure}
 \caption{Plot of the quantity $F(a)$ defined in the text for $Q=2$, $R=10^{5}$
and $\beta =0.348$.}
\label{IRbound}
\end{figure}

\begin{figure}
 \caption{K41-like static solution for the same value of parameters as Fig.\
\protect\ref{IRbound}.}
\label{ptfix}
\end{figure}

\begin{figure}
 \caption{ Scaling properties of $\Sigma_{n,p}$ for various $p$ and
$\beta =0.33$. }
\label{exscal}
\end{figure}

\begin{figure}
 \caption{Scaling exponents  $\sigma_{p}$ versus $p$ for $\beta =0.28$,
$\beta =0.33$, and $\beta=0.343$. For the lowest value of $\beta$ and $p\geq
3$, the  $\sigma_{p}$'s fall on a single straight line. There is more
 curvature for the other two values of $\beta$, indicating
multifractality. Nevertheless a linear growth is recovered for $p\geq 8$.}
\label{tp1}
\end{figure}

\begin{figure}
 \caption{A zoom on the evolution of the cross-over region with $\beta$. A
 particularity of the ON model compared to the GOY model is that it must get
 very close to the transition before showing a clear curvature of the
  $\sigma_{p}$'s. As a consequence, $\sigma_{p}$ is very close to
 zero for $p\leq 3$.}
\label{tp2}
\end{figure}

\begin{figure}
\caption{ Comparison of the statistics of the  ON and GOY models far
from the boundary between regular and chaotic dynamics.}
\label{GOYON1}
\end{figure}

\begin{figure}
\caption{Comparison of the statistics of the ON and GOY models in a regime
where multifractality is well established.}
\label{GOYON2}
\end{figure}

\begin{figure}
\caption{Amplitudes $b_{5}(\tau)$, $b_{12}(\tau)$, $b_{19}(\tau)$ within the
"projected" dynamics for $\beta$ =0.335 and the initial condition $b_{n}(0)=
\delta_{n,0}$. The emergence of a travelling wave is clearly demonstrated.}
\label{selfsim1}
\end{figure}

\begin{figure}
\caption{ Time evolution of the quantity $A(\tau)$ defined in Eq.\
(\protect\ref{defA}) of the text, for the same values of parameters as in Fig.\
\protect\ref{selfsim1}. The small oscillations of $A$ around its mean value
in the asymptotic regime are due to the discreteness of the lattice.}
\label{selfsim2}
\end{figure}

\begin{figure}
%\centerline{ \input{/home/CRTBT3/DOMBRE/SFP95/resi_1.tex} }
\caption{ Modulus and real part of residues for $\beta =0.39$, as obtained from
 an integration of the fictitious dynamics (\protect\ref{residyn}) with
$\delta =2$ and the initial condition $a_{n}=i\delta_{n,0}$ (N=29). }
\label{resi_1}
\end{figure}

\begin{figure}
 \caption{ Same as in Fig.\ \protect\ref{resi_1} but for $\beta =0.40$. In this
case the same pattern is always observed, independently of the initial
 condition. }
\label{resi_2}
\end{figure}

%--------------------Table1------------------------------------------------
\begin{table}
\begin{tabular}{lllll}
 \multicolumn{1}{c}{$\beta$} & \multicolumn{1}{c}{$z$} & \multicolumn{1}{c}{$
z-{2\over 3}$} & \multicolumn{1}{c}{$\gamma$} & \multicolumn{1}{c}{error} \\
\hline
0.15 & 0.996 & 0.329 & $\cdots$ & $\cdots$ \\
0.28 & 0.921 & 0.245 & 0.24 & $\pm 10^{-2}$ \\
0.33 & 0.889 & 0.223 & 0.213 & $\pm 10^{-2}$ \\
0.343 & 0.881 & 0.214 & 0.20 & $\pm 2 10^{-2}$ \\
0.348 & 0.878 & 0.212 & $\cdots$ & $\cdots$ \\ \hline
0.7 & 0.721 & 0.054 & \multicolumn{2}{c}{regular dynamics} \\
0.8 & 0.692 & 0.025 & \multicolumn{2}{c}{} \\
\end{tabular}
\caption{Exponents $z$ of self-similar solutions in the ON model for various
values of $\beta$. The left columns present
estimates obtained from Eq.\ (\protect\ref{computz}) after a numerical
 integration of Eq.\ (\protect\ref{projdyn}). The last digit is given with an
$\pm 1$ accuracy. The last two columns
present data extracted from statistical analysis. The comparison between
columns 3 and 4 show that scaling properties of self-similar solutions
account in a satisfying way for large order statistics in the chaotic part of
 the phase diagram, even close to the transition where $z$ remains rather big.}
\label{tab:1}
\end{table}
%
%
%
%--------------------Table 2 -----------------------------------------------
 \begin{table}
\begin{tabular}{lllll}
 \multicolumn{1}{c}{$\epsilon$} & \multicolumn{1}{c}{$z$} &
\multicolumn{1}{c}{$z-{2\over 3}$} & \multicolumn{1}{c}{$\gamma$} &
 \multicolumn{1}{c}{error} \\ \hline
0.398 & 0.684 & 0.018 & \multicolumn{2}{c}{regular dynamics} \\ \hline
0.5 & 0.719 & 0.052 & 0.12 & $\pm 3 10^{-2}$ \\
0.75 & 0.888 & 0.222 & 0.23  & $\pm 10^{-2}$ \\
0.8 & 0.946 & 0.279 & $\cdots$ & $\cdots$ \\
\end{tabular}
\caption{Same quantities as in table \protect\ref{tab:1} but for the GOY
 model. One observes that $z$ takes rather small values everywhere in the
 chaotic part of the phase diagram ($\epsilon > 0.398$). The disagreement
between columns 3 and 4 for $\epsilon =0.5$ is too large to be imputable to
a lack of statistics. }
\label{tab:2}
\end{table}

\appendix
\section*{}

In this Appendix we establish the equivalence between stable fixed points
of the dynamical system (\ref{residyn}) introduced in the Section 5 and
generic movable singularities in shell models.\\
First, we observe that static solutions of (\ref{residyn}), if they exist, are
 such that $\vec{N}[\vec{a}_{f}]=\lambda \vec{a}_{f}$ where the coefficient
of proportionality $\lambda =\displaystyle{\frac{\langle \vec{a}_{f},
\vec{N}[\vec{a}_{f}]\rangle }{\langle \vec{a}_{f}, \vec{a}_{f} \rangle }}$
 obeys~:
\begin{equation}
\lambda =Re(\lambda )+i\delta Im(\lambda )
\end{equation}
If $\delta \neq 1$, $\lambda$ is bound to be real. The case $\lambda =0$
corresponds to  $\vec{a}_{f}$ being an inertial fixed point
($\vec{N}[\vec{a}_{f}]= \vec{0}$). We discard this possibility (whose occurence
would be easily identified in practice) and
concentrate on the more interesting case of a finite value of $\lambda $.
Then $\vec{a}=-\displaystyle{\frac{\vec{a}_{f}}{\lambda }}$ verifies as it
 should $\vec{N}[\vec{a}]=-\vec{a}$. \\
Now we must ask about stability properties of $\vec{a}_{f}$. By linearizing
the system of differential equations (\ref{residyn}) around their fixed
point, one finds the following evolution of small perturbations
$\delta \vec{a}$~:
\begin{equation}
\frac{d}{dt} \delta \vec{a} = \lambda (M+1) \delta \vec{a} + \ldots
\end{equation}
where the terms hidden behind the dots are all directed in the direction
of $\vec{a}_{f}$ and have been omitted for simplicity. As in Eq.\
(\ref{equacorr}) of Section 5,  $M$ is the Jacobian matrix of first order
derivatives of the non linear kernel $\vec{N}$ evaluated at point $\vec{a}$
 defined above. It appears that the space ``transverse" to $\vec{a}_{f}$
 belongs as
a whole to the stable manifold of $\vec{a}_{f}$, if and only if all the
 eigenvalues $\mu_{i}$ of $M$ for $1\leq i \leq N$ have a real part
bigger (resp.\ smaller) than -1 with $\lambda$ negative (resp.\ positive).
The second possibility would lead to a divergence of the trace of $M$ in
the limit $N\rightarrow +\infty$, which contradicts the assumption of a
 finite norm for $\vec{a}$. We are thus left with $\lambda <0$ and by the
 same token N eigenvalues $\mu_{i}$ of real part bigger than -1, which is
 nothing but the criterion for genericity established in Section 5.\\
Finally, let us consider perturbations along the direction of $\vec{a}_{f}$
or $\vec{a}$. Since the dynamics (\ref{residyn}) preserves the norm of
 $\vec{a}(\tau )$, they reduce to phase fluctuations which may be
 parameterized as $\vec{a}(\tau ) = e^{i\theta (\tau )} \vec{a}_{f}$. The
 phase $\theta (\tau )$ is found to obey the equation of motion~:
\begin{equation}
 \frac{d}{d\tau } \theta =\lambda (\delta -1) \sin \theta
\end{equation}
Therefore complete stability of $\vec{a}_{f}$ requires $\delta > 1$ (since
$\lambda < 0$), as announced in the main text of Section 5.

 \end{document}